# Artificiality in Social Sciences

Jean-Philippe Rennard
Grenoble Graduate School of Business
Jp at rennard.org

**Abstract**: This text provides with an introduction to the modern approach of artificiality and simulation in social sciences. It presents the relationship between complexity and artificiality, before introducing the field of artificial societies which greatly benefited from the computer power fast increase, gifting social sciences with formalization and experimentation tools previously owned by "hard" sciences alone. It shows that as "a new way of doing social sciences", artificial societies should undoubtedly contribute to a renewed approach in the study of sociality and should play a significant part in the elaboration of original theories of social phenomena.

## Introduction

The "sciences of the artificial" deal with synthesized things which may imitate natural things; which have functions and goals and which are usually discussed in terms of imperatives as well as descriptives. Imitation with computer is now usually termed simulation and is used to understand the imitated system (Simon, 1996).
Artificiality has invaded science over the last thirty years and physicists, chemists or biologists now daily use widespread computing tools for simulations. Social sciences did not set this trend aside (Halpin, 1999). This chapter will first introduce the essential link between complexity and artificiality before presenting the highly promising field of artificial societies.

## Complexity and artificiality

Since the seminal book of Herbert Simon in 1969 (Simon, 1996), the sciences of the artificial knew a jerky evolution. In the field of artificial intelligence, the excessive ambitions of the sixties were considerably lowered in the seventies, before knowing a new wave of optimism in the mid eighties. The renewed interest toward artificiality originates in new approaches of artificial intelligence and in the success of the highly innovative related fields of artificial life (Langton, 1989) and artificial societies (Gilbert & Conte, 1995; Epstein & Axtell, 1996). Artificial life is at the crossroad of the rebirth of artificiality and offers lots of nice examples illustrating this revival, like this one:
Many ant species tend to form piles of corpses (cemetery) in order to clean their nest. Experiments with different species showed that if corpses are randomly distributed, ants tend to gather them in some clusters within few hours.
Deneubourg, Goss, Franks, et al. (1991) proposed a simple model of corpses gathering (see also Bonabeau, Dorigo, & Théraulaz, 1999). They designed virtual ants having the following behaviors:

- The probability for an ant to pick up a corpse is $p_p = (k_1/(k_1 + f))^2$ with $k_1$ a threshold constant and $f$ the fraction of perceived corpses in the neighborhood.

- The probability for an ant to deposit a corpse is: $p_d = (f/(k_2 + f))^2$ with $k_2$ a threshold constant. Deneubourg, Goss, Franks, et al. (1991) computed $f$ as the number of items perceived during the last $t$ periods divided by the largest number of items that can be encountered during the last $t$ periods.

To put it simply, virtual ants tend to pick-up isolated corpses to drop them in dense zones. The result (see figure 1) is close to the real phenomenon.

<< FIGURE 1 >>

**Figure 1. Virtual ant cemetery**

Highly simple virtual individuals ("agents") without any knowledge of the global process, manage to carry out cemetery building. Furthermore, it "suffices" to define different types of objects to obtain sorting capabilities, like for example larval sorting observed in anthills. The gathering or the sorting process *emerges* from the interactions of simple agents.

## *Emergence*

Emergence can be defined as the qualities or properties of a system which are new compared with the qualities or properties of the components isolated or differently organized (Morin, 1977). According to Gilbert (1995b): "Emergence occurs when interactions among objects at one level give rise to different types of objects at another level. More precisely, a phenomenon is emergent if it requires new categories to describe it that are not required to describe the behavior of the underlying components. For example, temperature is an emergent property of the motion of atoms. An individual atom has no temperature, but a collection of them does."

Most authors consider that emergence relies on three conditions:
1. The global process is distributed, there is no central control and the result depends on the interactions between components.
2. The process is autonomous, there is no external controller.
3. The process is not at the same *level* as the components. The language or concepts used to describe the emergent process are different from the language or concepts used to describe the components. The "Test of emergence" thus relies on the *surprise* engendered by the difference between the language $L_1$ used to design components, and the language $L_2$, used to describe the resulting process (Ronald, Sipper, & Capcarrère, 1999). According to Steels (1997), this change of language is sufficient to characterize emergence.

We can clarify this concept with the classical and very general formalization proposed by (Baas, 1994), which is based on three elements:
1. A set of *first order structures* $\{S_{i_1}^1\}, i_1 \in J_1$ with $J_1$ some index set finite or not. First order structures are primitive objects, abstract or physical; they can be organizations, machines as well as fields or concepts.
2. An *observational* mechanism $Obs$.
3. *Interactions Int*.

The new kind of structure resulting from the observed interactions of first order structures is: $S^2 = R(S_{i_1}^1, Obs^1, Int^1)_{i_1 \in J_1}$, where $R$ is the result of the interaction process and

$Obs^1 \equiv Obs(S^1_{i_1})$. Baas calls $S^2$ a *second-order structure*, and $\{S^2_{i_2}\}, i_2 \in J_2$ *families*.

The properties of the new unity resulting from the collection $S^2$ can be measured with an observational mechanism $Obs^2$. Then $P$ is an *emergent property* of $S^2$ iff

$$P \in Obs^2(S^2), \text{ but } P \notin Obs^2(S^1_{i_1}) \text{ for all } i_1.$$

The property $P$ belongs to the emergent structure $S^2$, but is absent from the components.

The $S^2$'s can interact to form a third order structure, and so on. A *N-th order structure* is then:

$$S^N = R(S^{N-1}_{i_{N-1}}, Obs^{N-1}, Int^{N-1}), i_{N-1} \in J_{N-1}$$

Baas calls it a *hyperstructure* and he considers that "*complexity often takes the form of a hyperstructure.*" (Baas, 1994, p.525, original italics). According to Simon (1996), hierarchies are necessary to allow the evolution of complex structures.

Baas distinguishes two different types of emergence:

- *Deductible or computable emergence*: A process or theory $D$ exists which allows to determine $P \in Obs^2(S^2)$ from $(S^1_{i_1}, Obs^1, Int^1)$. That is typically the case of engineering constructions or "trivial emergence" like temperature evoked above.
- *Observational emergence*: the emerging property $P$ cannot be deduced (e.g. consequences of Gödel's theorem).

Bedau (1997) considers less drastically, that *weak emergence* characterizes emerging properties that can only be derived by simulation. Most of the recent modeling works deal with this type of weak emergence (Chalmers, 2002).

Despite the thousands of pages published on emergence, or the recent emphasis on the *reduction principle* (the macrobehavior is reducible to the interactions of the components), (e.g. Holland, 1999; Kubik, 2003)), we are still far from an ontological concept of emergence (Emmeche, Koppe, & Stjernfelt, 1997), but, considering its success, emergence is undoubtedly epistemologically a fertile concept.

## *Bottom-up modeling*

Emergence is a key feature of those famous *non-linear systems* which are said to be *more than the sum of their parts* (Waldrop, 1992). Non-linear systems do not obey the *superposition principle* —the linear combination of solutions is not a solution. Their dynamic cannot be reduced to the simple (linear) combination of their components ones.

We have known since at least the end of the nineteenth century and Henri Poincaré (Poincaré, 1892), that the dynamic of such complex systems is unpredictable. The only way to know their state at a given step is to compute each step. The usual analytical method is of few help; the necessary mathematics are still to be invented. Even the small body of mathematics which directly deals with non-linearity depends upon linear approximations (Holland, 1999). Non-linearity thus challenges the traditional approach which tries to understand a system by analyzing its components: "The key feature of non-linear systems is that their primary behaviors of interest are properties of the *interactions between parts*, rather than being properties of the parts themselves, and these interactions-based properties necessarily disappear when the parts are studied independently." (Langton, 1989, p.41, original italics).

How to deal with emergence? How to study processes which are "more than the sum of their parts"? How to analyze properties that cannot be forecasted? The solution proposed by computer scientists is termed *bottom-up modeling*.

Bottom-up modeling is a very new way of building artificial systems. Since core properties disappear when the components are studied independently, bottom-up modeling is based on the gathering of interacting components. Corpses clustering or larval sorting models are then based on the building of rather simple *agents* (see below) which interact both with one another and with the environment. Such constructions and the study of the dynamic resulting from non-linear interactions of the simple components constitute the "bottom-up method". Instead of modeling the global dynamic of the studied system ("top-down method" usually based on differential equations) one merely models the components to study the potentially emerging regularities.

This *synthetic method* is at the heart of the revival of artificiality. Commenting the first workshop on artificial life, C. Langton stated: "I think that many of us went away […] with a very similar vision, strongly based on themes such as *bottom-up* rather than *top-down* modeling, *local* rather than *global* control, *simple* rather than *complex* specifications, *emergent* rather than *prespecified* behavior, *population* rather than *individual* simulation, and so forth." (Langton, 1989, p.xvi, original italics).

The $19^{th}$ century ended with Poincaré's discovery of the limits of the analytical method faced with non-linear systems. The $20^{th}$ century ended with the unprecedented quick spread of a machine able to deal with these systems. Computers are in fact surprisingly adapted to the analysis of non-linear systems. Besides their ability to iteratively compute equations which do not have analytical solutions, computers —particularly since the development of object oriented programming— can easily deal with populations of interacting agents, so contributing to the study of Bedau's weak emergence. "(…) Computer-based models offer a halfway house between theory and experiment [(…) and computer-based non-linear modeling] will certainly improve our understanding of emergence." (Holland, 1999, p.232).

Bottom-up modeling is based on the interactions of (usually) simple virtual individuals. It massively uses *multi-agent systems* (MAS).

## *Multi-Agent Systems*

MASs originate in *Distributed Artificial Intelligence* (DAI) and in artificial life. The basic idea of DAI is that intelligence is not only a matter of phenotype (brain) but also depends on the interactions with other individuals. Intelligence has a "social dimension" (Drogoul, 2005). The emergence of DAI is directly linked to the limits of the traditional symbolic AI (GOFAI) which tries to embed intelligence in a unique entity. The *cognitive school* of DAI associates a few complex agents to obtain some kind of group expertise (see e.g. Demazeau & Müller, 1991). The *reactive* school of DAI is more original. Strongly rooted in artificial life, it uses the insect (and animal) societies metaphor to try to obtain *emergent intelligent behaviors* by associating simple ("sub-cognitive") agents (Steels, 1990; Deneubourg, Goss, Beckers, & Sandini, 1991).

We have seen that cemetery building was modeled with "virtual insects" i.e. some software processes that imitates insects' behaviors. These virtual insects are *agents*. Jacques Ferber, one of the founders of the field, considers that an agent is a physical or virtual entity (Ferber, 1999):

- capable of acting.
- capable of communicating with other agents.
- driven by a set of tendencies. Autonomous agents act according to their own goals.
- having its own resources; but these resources depend on the environment. Agents are then open systems since they find resources in the environment, and close system, since they manage the use of these resources.
- having a partial representation of their environment. An agent thus do not have to "fully understand" its environment; above all it does not have to perceive the global result of its actions.

- possessing skills.
- possibly able to reproduce itself.
- tending to act according to its objectives.

"The agent is thus a kind of 'living organism', whose behavior, which can be summarized as communicating, acting and perhaps, reproducing, is aimed at satisfying its needs and attaining its objectives, on the basis of all the other elements (perception, representation, action, communication and resource) which are available to it." (Ferber, 1999, p.10).

Ferber's definition is restrictive and one can limit the characterization of agents to the following core properties (Wooldridge & Jennings, 1995):

- *autonomy*: agents operate according to their own control.
- *social ability*: agents can interact with one another through some kind of language.
- *reactivity*: agents can perceive their environment and react according to its change.
- *pro-activness*: agents act according to their own goals.

Figure 2 summarizes the structure of an agent.

<< FIGURE 2 >>

**Figure 2. An agent in its environment**

Bottom-up modeling uses interacting agents by building *multi-agent systems* (MAS). A MAS contains the following elements (Ferber, 1999): An environment $E$; a set of objects $O$ having a specific position in the environment; a set of agents $A$ with $A \subseteq O$; a set of relations $R$ linking the objects to each other; a set of operations $Op$ allowing the agent to "perceive, produce, consume, transform and manipulate" objects; operators able to apply the operations and to process the reaction of the environment. MASs and *Agent Based Modeling* (ABM) are the base of social simulation (see e.g. the *Iterated Prisoners Dilemma*—IPD (Axelrod, 1984, 1997)) and artificial societies (Conte, Gilbert, & Sichman, 1998).

# Artificial Societies

How to connect virtual agents with human societies? Humans are quite different from ants and despite real progress —thanks to the quick growth of computer power— the intelligence of the most sophisticated agent ever programmed cannot be compared to human intelligence. The 2005 Nobel Prize in Economics was attributed to Thomas C. Schelling (along with Robert J. Aumann) who proposed in 1971 (Schelling, 1971, 1978) a far ahead of one's time experiment, which will help us understand the link between agents and human societies.

*The seminal model of Thomas Schelling*

Schelling wanted to understand the pre-eminence of geographical segregation between black and white in American cities despite the fact that when they are questioned, citizens refute any desire of segregation. He designed very simple agents of two distinct colors ("black and white"), having the following abilities:

- Each agent can compute the fraction of neighbors having the same color.
- If this fraction is below the agent preference, then the agent moves to an unoccupied place which satisfies its preference.

Schelling used cellular automata to implement its experiment. Very briefly, cellular automata are lattice of sites whose states —belonging to a finite set— evolve in discrete time step according to rules depending on the states of the neighbors sites. In a two dimensions implementation, Schelling used a "Moore" neighborhood, i.e. neighbors are the eight closest squares. The rules were:

- If an agent has two neighbors, it will not move if at least one is of the same color.
- If an agent has three to five neighbors, it will not move if at least two are of the same color.
- If an agent has six to eight neighbors, it will not move if at least three are of the same color.

These rules are compatible with a fully integrated structure. The initial state of Schelling (see figure 3) is thus an attractor since no agent needs to move. Schelling showed that a slight perturbation of this initial state is sufficient to give rise to a dynamic quite inevitably leading to segregation (see figure 3).

<< FIGURE 3 >>

**Figure 3. Schelling's model**

Schelling's model clearly demonstrates that local interactions (*micromotives*) lead to global structures (*macrobehavior*, (Schelling, 1978)). More important, he showed that the macrobehavior can be different from the underlying micromotives, since segregation occurs even when preference rules are compatible with integrated structure. Nowak and Latané (1993) used an extended model to study *Dynamic Social Impact* i.e. the change of attitudes or beliefs resulting from the action of other individuals. They notably showed that the system achieved stable diversity. The minority survived, thanks to a clustering process of attitudes, not because individuals moved, but due to the attitude change process. (Latané, 1996). The observed macrobehaviors are very robust. Schelling's and Latané's models were tested under a wide range of parameters and quite always evolve towards the same type of attractors. Pancs and Vriend (2003) recently enlarged the study of segregation process showing that it tends to occur even if people are anxious that segregation should not occur.

Both these examples show that some complex social dynamics can be modeled from simple basis: "(…) there is a spirit in the air which suggests that we should look for simple explanations of apparent complexity." (Gilbert, 1995b). Stephen Wolfram recently brought a strong justification to this quest for simplicity (Wolfram, 2002). Its *Principle of Computational Equivalence* states that: "(…) almost all processes that are not obviously simple can be viewed as computations of equivalent sophistication. (…) So this implies that from a computational point of view even systems with quite different underlying structures (…) can always exhibit the same level of computational sophistication. (…) And what it suggests is that a fundamental unity exists across a vast range of processes in nature and elsewhere: despite all their detailed differences every process can be viewed as corresponding to a computation that is ultimately equivalent in its sophistication." (Wolfram, 2002, pp.717-719). Without going as far as Wolfram, it is now clear that at least some social phenomena can be modeled with interacting sub-cognitive agents.

- The Newtonian model uses systems of differential equations to study equilibrium; the best example being equilibrium theory in economics —which is also a brilliant example of the consequences of oversimplification motivated by the will to obtain tractable equations; the results having few to do with reality.
- Considering the difficulty to write the equations of the system, the statistical model tries to discover regularities; the best example being the study of *"social forces"* by Durkheim in 1897 (Durkheim, 2004).

Schelling's or Latané's models are then quite a new way of doing social sciences based on virtual experiments inside artificial societies.

## *Artificial Societies as a new way of doing social sciences*

The field of artificial societies is based on the strong assumption that human societies are complex systems (Goldspink, 2000). Analysis is unable to point the source of macro-properties since there is no localized source, but a distributed process which obliges to consider the system as a whole (Goldspink, 2002). Furthermore, they are *complex adaptive systems* (CAS) i.e. systems where agents can learn and modify their rules according to their previous success (that is of course also the case of animal or insect societies, but the specificity of human —cognitive— societies is that they can also learn from their failures). Schelling's segregation process or Nowak and Latané's clustering process of people sharing the same opinion are emergences or "regularities at the global level" (Gilbert, 1995a). "As the number of elements and interactions of a system is increased, we can observe an *emergent complexity*. But *somehow*, regularities arise and we can observe *emergent simplicity*" (Gershenson, 2002, original italics).

Artificial societies then try to obtain emergent regularities: "(…) the defining feature of an artificial society model is precisely that *fundamental social structures and group behaviors emerge from the interaction of individual agents operating on artificial environments* (…)."(Epstein & Axtell, 1996, p.6, original italics). Considering European contributions to social modeling, Gilbert wrote: "One of the major objectives of the approach being reviewed here is to generate through simulation, emergent phenomena and thus to understand and explain the observable macro-level characteristics of societies." (Gilbert, 2000).

This is quite a new way of doing science; so new that simulation is said to be "a third way of doing sciences" (Axelrod, 2006) different from deduction and from induction. In the fields of artificial intelligence and artificial life, Luc Steels termed it the *synthetic method* (see figure 4) (Steels & Brook, 1994).

<< FIGURE 4 >>

**Figure 4. Inductive vs. synthetic method**

Induction starts from observed facts and uses inferences to build a theory potentially able to globally explain the observed facts. The theory is then validated through the test of predicted facts. The synthetic method starts like induction from the observed facts and the inferred theory (but it can also start like deduction from a set of assumptions). On this basis, the synthetic method engineers an artificial system, the objective being that, while operating, this system will behave like the real one, thus confirming the tested theory.

In their seminal work, Epstein and Axtell (1996) considered that artificial societies models may change the way we think about *explanation* in the social sciences. "Clearly, agent-based social science does not seem to be either deductive or inductive in the usual senses. But then what is it? We think *generative* is an appropriate term. The aim is to provide initial microspecifications that are *sufficient to generate* the macrostructures of interest." (Epstein &

Axtell, 1996, p.177). This generative interpretation is directly linked to the disjunction between determinism and predictability which is a huge epistemological consequence of complexity sciences. Even if we perfectly understand the concerned forces, we are unable to predict the evolution of the system (Croquette, 1997).

*A high potential to stimulate novelty*

Agent based modeling is potentially a highly powerful tool for social scientists. Axelrod and Tesfatsion (forthcoming) recently synthesized its goals with four forms:
- *Empirical understanding*: why have regularities emerged?
- *Normative understanding*: how can models help to define the good norms/design? How to know if a given decision is positive for the society?
- *Heuristic*: How to attain greater insight about fundamental mechanisms in social systems?
- *Methodological advancement*: How to give researchers the method and tools to rigorously study social systems?

Practically, these four forms rely on three pillars: Formalization, experiments and ability to study the *macro to micro problem*.

*Formalization*

Apart from the verbal and mathematical symbol systems, computer simulation can be considered as the "third symbol system" (Ostrom, 1988). Any theory originating in the first two models can be expressed in the third one. Simulation can then be considered as formal models of theories (Sawyer, 2004). That is an important point since computer symbols are more adapted to social sciences than mathematical ones (Gilbert & Troitzsch, 2005, pp.5-6):
- Programming languages are more expressive and less abstract than mathematical techniques.
- Programs deal more easily with parallel processing.
- Programs are modular. Major changes can easily be made, that is not the case of mathematical systems.

Computer modeling thus helps social scientists to formalize their theories. The difficulty — not to say the impossibility— to mathematically formalize many social sciences theories is considered to be a great weakness by "hard" scientists. This inability is closely linked to the inability of mathematics to deal with distributed emergent processes. Computer modeling can thus contribute to give social sciences some of the scientific tools they need to rigorously express their theoretical models.

*Experiments*

Simulation can be considered as a new experimental methodology. Gilbert and Conte (1995) defined it as "exploratory simulation". Such explorations can contribute to social sciences notably in the following ways:
- Modeling allows a *culture-dish methodology*. The modeler designs the agents and the initial state of its society and studies its temporal evolution (Tesfatsion, 2002). Any sort of experiments can be carried out since the modeler has a complete control on the model. It is then possible to study the consequences of any given modification. This will notably contribute to the analysis of the minimal set of parameters and system characteristics necessary to give rise to a given behavior as well as to the analysis of the attractors of dynamic social systems (Goldspink, 2002). The ability to carry out experiments is something very new for social scientist that usually cannot test their theory in the field. Like formalization this contributes to bring closer social and "hard" sciences methods.

- Modeling is potentially able to contribute to original discoveries. The same way the classification of cellular automata permitted to propose an original analysis of complex systems (Wolfram, 1984; Langton, 1990), simulations can play a role in the discovery of general, yet unattainable, laws. Implicit unknown effects can be detected (Gilbert & Conte, 1995). This ability to stimulate discovery does not only stand on the possibility to carry out otherwise impossible experiments, but also on the capacity of emergent modeling to give rise to original cognitive processes. In the field of artificial life, Cariani (1992) emphasizing non-stochastic models like the Game of Life, pointed out the fact that emergence relies on a cognitive process; a process is emergent only according to its observer: "*The interesting emergent events that involve artificial life simulations reside not in the simulations themselves, but in the way that they change the way we think and interact with the world.* Rather than emergent devices on their own right, these computer simulations are catalyst for emergent processes in our minds; they help us create new ways of seeing the world." (Cariani, 1992, p.790, original italics).

- Modeling can go beyond some of the limits of the statistical tools usually used by social scientists, e.g. qualitative changes can be analyzed through simulation (see (Pyka, 2006)). Simulation also helps the study of processes. Usual statistical analyses study the correlations between variables at a given time. Simulations embed the processes which lead to these correlations (Gilbert & Troitzsch, 2005). Since social systems are fundamentally dynamic, simulation allows formalizing processes beyond the scope of statistical analysis. Furthermore, statistic is based on linearity assumptions which oblige to over simplify the observed facts. Simulation does not suffer from this limit.

- Modeling is not concerned by the technical limits of mathematical formalization. For example, mathematical formalization obliges to consider agents as equivalent whereas simulation is able to manage heterogeneous population. In the same vein, simulation allows to relax assumptions necessary to obtain tractable equations (e.g. the rationality of economic agents). In economics, the highly promising field of Agent-Based Computational Economics (ACE) (Tesfatsion, 2002) clearly illustrates the potential of simulations.

- More generally, the same way artificial life allows the study of "Life as it could be" (Langton, 1989), artificial societies allow the study of "Societies as they could be" (Gilbert, 2000), thus giving social sciences an unprecedented tool to understand fundamental invariants (Rennard, 2004).

*Study of the macro to micro problem*

The *macro to micro problem*—how to describe the relationship between macro-phenomena characterizing the dynamic of a system as a whole and micro-phenomena characterizing the dynamic of the components of the system— is a central issue of social sciences, but also of DAI (Schillo, Fischer, & Klein, 2000).
Simulation is a ground-breaking tool to study the core problem of the micro/macro relations. The relations between different levels (individual, organization, societal) and the potential associated lock-in can be studied. Artificial life with its widely studied concept of *Dynamical hierarchy* which "refers to a system that consists of multiple levels of organization having dynamics within and between the entities described at each of the different levels." (Lenaerts, Chu, & Watson, 2005, p.403), should contribute to this study. Simulation can be used to study both the *micro to macro* and the *macro to micro problems* (Sawyer, 2003). Schelling's or

Latané's models thus show how regularities can arise from micro-interactions. But such models also show that these regularities then constraint the system and impact the behaviors of individual agents. More directly, it is possible to conceive simulations that specifically study the impact of macro-phenomena. For example, Axtell (2000) while studying retirement behaviors, showed that modifying the sole network connections between agents can lead to great changes of the overall society behavior.

The study of the micro/macro problem through simulation remains nevertheless very difficult while studying societies. In fact, humans are not limited to basic behavior, they notably have the ability to grasp macro-level phenomena and they can adjust their behavior according to this. That is what Gilbert (2000) terms *second order emergence*, characterizing systems where agents can detect and react to emergent properties. Models should then embed both the emergence of macro-properties and the ability to deal with the effects of these macro-properties on self-aware individuals. This remains a challenge (Gilbert, 1995b).

*Limits*

Artificial societies is a very recent field in which huge problems still are to be solved that challenges these researches.

A first set of problems relies on the cognitive dimension of human societies. Guided by the success of artificial life, many artificial societies are based on reactive DAI, one of the most famous example being the *Sugarscape* of Epstein and Axtell (1996). The complexity of human cognition has a deep impact on the structuring of societies.
- Self-awareness and the related second order emergence should be modeled.
- Interpretativism in sociology leads to the idea that meanings are parts of the actions. "(…) meanings and concepts describing both the physical and the social world are said to be socially constructed by members of society" (Gilbert, 2000). Simulations should then embed the corresponding social constructions.

As a consequence, artificial societies must find a way to associate cognitive and reactive DAI. This remains both a theoretical (how to build cognitive agents) and a practical (how to have sufficient computing power) problem.

A second set of problems is linked to the tools and methods used for modeling and simulation. First of all, simulation uses tools that may make implicit assumptions having nothing to do with the tested theory. For example, the use of cellular automata assumes that the world is a regular grid, which may have massive consequences on the global dynamic of the simulation (Troitzsch, 1997). Then simulation tends to develop its own finality, hence the importance to ground it in social theories in order to avoid the trend to develop simulations for themselves and to mistake them for reality. The balance is difficult to find: "If our 'toy models' serve only to reify and naturalize the conventional social science wisdom, then they are indeed a Medusan mirror, freezing the victim by the monster's glance" Lansing (2002, p.289).

The gap between social sciences and computer sciences also challenges the field. Some social sciences theories are mainly descriptive and discursive and such approaches may be very difficult to formalize through simulation. Moreover, despite common issues, the discussion between computer scientists and social scientists remains very difficult. For computer scientists, non formalized discursive social theories often seem blurred and they have difficulties in understanding them. Social scientists are often reluctant facing computer programming and they usually consider that computer scientists do not understand the complexity of human societies.

Finally, the core problem (which is not limited to artificial societies) of "how to obtain from local design and programming, and from local actions, interests, and views, *some desirable and relatively predictable/stable emergent results*" (Castelfranchi, 2000, original italics) still remains to be solved.

# Conclusion

The field of artificial societies, despite old roots, is now only ten years old. Along with artificial life, it participates to an emerging way of doing science. This way still has to reach maturity, but will undoubtedly contribute to complement more traditional methods. The debate now is not to choose between usual methods and methods originating in artificiality, but to convince "traditional" scientist that artificiality is not limited to some kind of, possibly funny, computer game and to find ways of building stronger bridges between these practices of science. The growing easiness of computer programming and the quick spread of computer culture among young scientists is potentially a promise of quick evolution of artificiality in social sciences; no doubt this will contribute to renew the field.

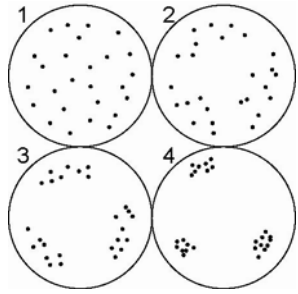

**Figure 1. Virtual ant cemetery**

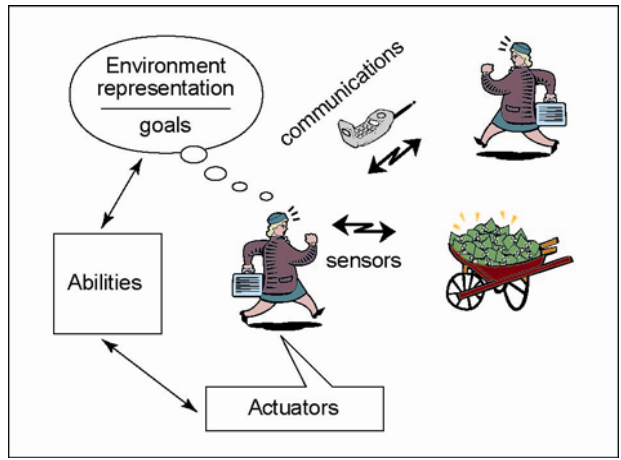

**Figure 2. An agent in its environment**

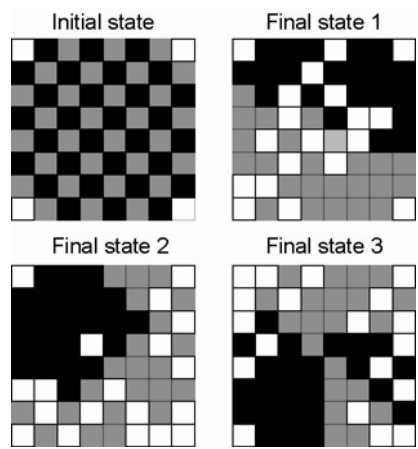

**Figure 3. Schelling's model**

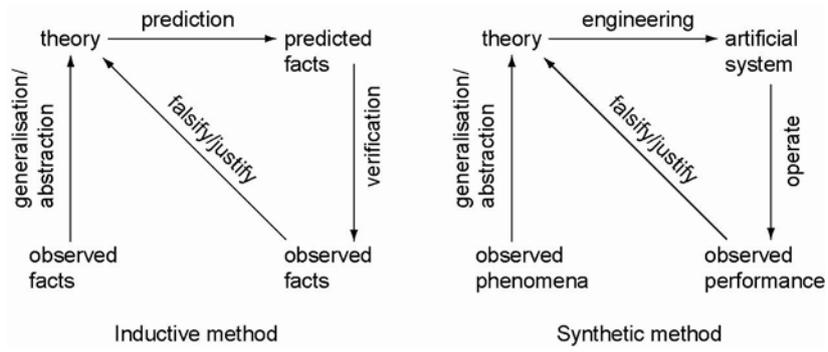

Adapted from (Steels & Brook, 1994).

**Figure 4. Inductive vs. synthetic method**